# Scaling Laws for Planetary Dynamos.

P. A. Davidson
Dept. Engineering, University of Cambridge, U.K.,
19th February 2013, Submitted to *Geophys J. Int.*

**Abstract**. We propose new scaling laws for the properties of planetary dynamos. In particular, the Rossby number, the magnetic Reynolds number, the ratio of magnetic to kinetic energy, the Ohmic dissipation timescale and the characteristic aspect ratio of the columnar convection cells are all predicted to be power-law functions of two observable quantities: the magnetic dipole moment and the planetary rotation rate. The resulting scaling laws constitute a somewhat modified version of the scalings proposed in Christensen & Aubert (2006) and Christensen (2010). The main difference is that, in view of the small value of the Rossby number in planetary cores, we insist that the nonlinear inertial term, $\mathbf{u} \cdot \nabla \mathbf{u}$, is negligible. This changes the exponents in the power-laws which relate the various properties of the fluid dynamo to the planetary dipole moment and rotation rate. Our scaling laws are consistent with the available numerical evidence.

## 1. Introduction

In table 1 we tabulate some of the properties of those terrestrial planets, moons and gas giants that are thought to have active dynamos. Much of the data in the table is adapted from de Pater & Lissauer (2001), updated where appropriate. The mean axial magnetic fields in the conducting cores and in the planets are calculated using the well-known relationship (see for example, Jackson, 1998),

$$\overline{B}_z = \frac{2\mu}{3V}|\mathbf{m}|, \qquad (1)$$

where $B_z$ is the axial component of the magnetic field, $\overline{B}_z$ its volume average, $\mathbf{m}$ the planetary dipole moment, $\mu$ the permeability of free space, and $V$ the volume of the planet or conducting core, as appropriate. We have not included the ice giants in table 1 as their complex magnetic field structure and large dipole inclination suggests that the origin of their magnetic fields is quite different to that of the other planets.

| Planet | Rotation period (days) | Mean planetary radius ($10^3$km) | Core radius, $R_C$ ($10^3$km) | Dipole moment, **m** ($10^{22}$Am$^2$) | Mean $\overline{B}_z$ in core (Gauss) | Mean $\overline{B}_z$ in planet (Gauss) |
|---|---|---|---|---|---|---|
| Mercury | 58.6 | 2.44 | 1.8 | 0.004 | 0.014 | 0.006 |
| Earth | 1 | 6.37 | 3.49 | 7.9 | 3.7 | 0.61 |
| Ganymede | 7.15 | 2.63 | ~0.5 | ~0.013 | ~2 | ~0.01 |
| Jupiter | 0.413 | 69.5 | 55 | 150,000 | 18 | 9 |
| Saturn | 0.444 | 58.1 | 29 | 4500 | 3.7 | 0.5 |

Table 1 Estimated properties of those terrestrial planets and gas giants that are thought to have dynamos.

It is immediately apparent that Mercury has a very weak magnetic field (about 500 times weaker than that of the Earth) with the Elsässer number, $\Lambda = \sigma B_z^2/\rho\Omega$, being very much less than unity, as indicated in table 2. (Here $\sigma$ is the electrical conductivity, $\rho$ the fluid density and $\Omega$ the planetary rotation rate.) The reason why Mercury's field is so weak is disputed. Some suggest that, based on thermal evolution models of Mercury, the convection in the core is strongly suppressed by stable stratification, with convection confined to the deepest parts of the core. This theory also accounts for the strongly dipolar structure of Mercury's field, as higher harmonics would be shielded by the outer regions of the conducting core (see, for example, Christensen, Schmitt & Rempel, 2009). In any event, what sets Mercury apart from the dynamos operating in the Earth and in the gas giants is the uniquely low value of the Elsässer number. In all other cases we find $\Lambda = O(10^{-1})$, possibly hinting at an order-of-magnitude balance between the Lorentz and Coriolis forces (sometimes called the strong-field regime). By contrast, Mercury, with its low Elsässer number, is usually considered to operate in a weak-field regime (see, for example, Rüdiger & Hollerbach, 2004), which is quite different to the other dynamos.

It is important to note, however, that there are other ways in which the mean core magnetic field may be made dimensionless, and depending on the dimensionless grouping one uses, Mercury's field may or may not look anomalous. Compare, for example, the third and fourth columns in table 2. Clearly, a more careful analysis is required before conclusions can be drawn about Mercury.

| Planet | Mean $B_z$ in core (Gauss) | Elsässer number $\Lambda = \sigma B_z^2 / \rho \Omega$ | Scaled magnetic field $\dfrac{B_z / \sqrt{\rho \mu}}{\Omega R_C}$ |
|---|---|---|---|
| Mercury | 0.014 | $6 \times 10^{-5}$ | $5.5 \times 10^{-6}$ |
| Earth | 3.7 | 0.08 | $1.3 \times 10^{-5}$ |
| Ganymede | ~2 | ~0.2 | ~$3.6 \times 10^{-4}$ |
| Jupiter | 18 | 0.5 | $5.2 \times 10^{-6}$ |
| Saturn | 3.7 | 0.02 | $2.2 \times 10^{-6}$ |

Table 2 Characteristic values of the Elsässer number, $\Lambda = \sigma B_z^2 / \rho \Omega$, and of the scaled magnetic field, $B_z / \Omega R_C \sqrt{\rho \mu}$, based on the estimated mean axial field strength in the core. We use the crude estimates of $\lambda \sim 2 \, \text{m}^2/\text{s}$ and $\rho \sim 10^4 \, \text{kg/m}^3$ for the terrestrial planets, and $\lambda \sim 30 \, \text{m}^2/\text{s}$ and $\rho \sim 10^3 \, \text{kg/m}^3$ for the gas giants, though estimates of $\lambda$ for Saturn vary considerably.

There has been a long tradition of trying to relate the dipole moments of the various planets, or equivalently their mean internal axial fields, to the observable properties of those planets, and many of these scaling theories are discussed in the review by Christensen (2010). An early suggestion was that $\Lambda = \sigma B_z^2 / \rho \Omega \sim 1$ for all planetary dynamos, or in terms of the magnetic energy per unit mass, $B_z^2 / \rho \mu \sim \lambda \Omega$, where $\lambda$ is the magnetic diffusivity. Leaving aside Mercury's particularly weak magnetic field, table 2 suggests that there may be some merit to this hypothesis, though there is not much variation in rotation rate between the Earth, Jupiter and Saturn, so this is not a very stringent test.

An alternative and largely successful approach is that of Christensen & Aubert (2006), Christensen, Holzwarth & Reiners (2009) and Christensen (2010). The hallmark of Christensen's scaling laws is the assertion that $\Omega$, while a crucial catalyst for dynamo action, is ultimately unimportant when it comes to determining the magnitude of the saturated, steady-on-average magnetic field. Rather, the saturated magnetic energy density, $B^2 / \rho \mu$, should be determined by the rate of working of the buoyancy force which drives the dynamo. There is solid support for this assertion from numerical dynamo simulations (see Christensen, 2010, and references therein) and so we embrace this hypothesis, using it to full advantage in the analysis that follows.

Our starting point, however, is somewhat different. It is an empirical observation that the mean axial vorticity in rapidly-rotating turbulent convection is independent of the background rotation rate. As we shall see, this is equivalent to Christensen's suggestion that the magnetic energy density in saturated dynamos is determined by the rate of working of the buoyancy force and independent of $\Omega$ (except to the extend that $\Omega$ may influence the rate of working of the buoyancy force). Starting from this single assumption we deduce, from first principles, a number of scaling laws which are shown to be consistent with the available numerical and observational evidence.

## 2. Revisiting the Hypothesis that the Magnetic Energy is Determined by the Rate of Working of the Buoyancy Forces.

There have been many attempts to arrive at a scaling law for the magnetic energy density in planetary dynamos. Typically, one starts with an assumed force balance in the core, which in turn leads to scaling laws for the characteristic velocity and for $B^2/\rho\mu$. However, in some cases the assumed force balance has the nonlinear inertial term, $\mathbf{u}\cdot\nabla\mathbf{u}$, as an order-one quantity (see the review by Christensen, 2010), which is an assumption we wish to avoid. Never-the-less, as noted above, there is strong numerical support for the hypothesis that the saturated magnetic energy density in planetary dynamos is simply determined by the rate of working on the buoyancy force and is independent of $\Omega$ (except to the extent that $\Omega$ may influence the rate of production of energy by the buoyancy forces). Consequently, we take a rather different approach here and decouple the question as to whether or not $B^2/\rho\mu$ is independent of $\Omega$ from any assumed velocity scaling in the core. Let us start by considering why $B^2/\rho\mu$ might be independent of $\Omega$.

Consider the momentum equation written in the rotating frame of reference and for a Boussinesq fluid which is convecting under the influence of a temperature field $T$:

$$\frac{\partial \mathbf{u}}{\partial t} = 2\mathbf{u}\times\mathbf{\Omega} - \nabla(p/\rho) - \beta T'\mathbf{g} + \rho^{-1}\mathbf{J}\times\mathbf{B} + \nu\nabla^2\mathbf{u} \ . \qquad (2)$$

Here $\mathbf{u}$ is the velocity field, $p$ the pressure, $\nu$ the viscosity, $\mathbf{g}$ the gravitational acceleration, $\beta$ the thermal expansion coefficient, and $T' = T - T_0$ where $T_0$ is a reference temperature. Note that the Rossby number, $\mathrm{Ro} \sim u/\Omega R_C$, $R_C$ being the core radius, is usually estimated to be extremely small in planetary cores, perhaps around $\mathrm{Ro} \sim 10^{-6}$, which obliges us to drop the nonlinear inertial term $\mathbf{u}\cdot\nabla\mathbf{u}$ in (2). Taking the dot product with $\mathbf{u}$ we have the energy equation

$$\frac{\partial}{\partial t}\left(\mathbf{u}^2/2\right) = -\nabla\cdot(p\mathbf{u}/\rho) - \beta T'\mathbf{u}\cdot\mathbf{g} + \rho^{-1}(\mathbf{J}\times\mathbf{B})\cdot\mathbf{u} - \varepsilon \ ,$$

where $\varepsilon$ is the viscous dissipation rate. When combined with the magnetic induction equation we obtain,

$$\frac{\partial}{\partial t}\left(\frac{\mathbf{B}^2}{2\rho\mu} + \frac{\mathbf{u}^2}{2}\right) = -\nabla\cdot\left[(p\mathbf{u}/\rho) + (\mathbf{E}\times\mathbf{B}/\rho\mu)\right] - \beta T'\mathbf{u}\cdot\mathbf{g} - \frac{\mathbf{J}^2}{\rho\sigma} - \varepsilon \ , \qquad (3)$$

where the extra divergence incorporates the Poynting flux, $\mathbf{E}\times\mathbf{B}/\mu$. Note that $T'$ and $\mathbf{u}$ may contain both steady and fluctuating contributions, corresponding to a combination of steady-on-average convection and turbulence. Note also that a similar energy equation can be derived for a non-Boussinesq fluid.

We shall focus here on low-$\Pr_m$ fluids, $\Pr_m = \nu/\lambda \ll 1$, as encountered in planetary dynamos. (It should be noted, however, that many of the numerical simulations correspond to $\Pr_m \geq 1$, and so care must be taken when comparing our predictions with those simulations.) For a turbulent dynamo at low magnetic Prandtl number we expect the destruction of energy by the Joule dissipation to dominate over the viscous dissipation, and so the total energy ceases to grow when the rate of working of the buoyancy forces is balanced by the Joule dissipation. Thus, when the dynamo is steady-on-average, integrating over all space and taking a time average yields the energy balance

$$\left| \int_{R_C} \beta \overline{T'\mathbf{u}} \cdot \mathbf{g} dV \right| = \int_{R_C} \overline{\mathbf{J}^2}/\rho\sigma \, dV \sim \frac{\mathbf{J}^2}{\rho\sigma} V_C \sim \frac{B^2}{\rho\mu} \frac{\lambda}{\ell_{\min}^2} V_C, \qquad (4)$$

[turbulent, low-$\Pr_m$ dynamo]

where the over-bar indicates a time average and $\ell_{\min}$ is the *smallest characteristic length scale* associated with the magnetic field. (Since $\Pr_m$ is assumed to be small there may, of course, be smaller length scales associated with the velocity field.) An important point to note about (4) is that the right-hand side is dominated by the small scales in the turbulence; in effect the right-hand side defines $\ell_{\min}$, while the left provides a means of evaluating it in terms of the rate of production of energy. However, the force balance in the core is dominated by the large scales which contribute to $\mathbf{B}$, $\mathbf{u}$ and $T'$, and so it is important to distinguish between the *micro-scale* $\ell_{\min}$ and the *integral scales* of the motion which typically enter into the force balance.

Now we might expect $\ell_{\min}$ to be determined by the balance between convection and magnetic diffusion, and so satisfy $u_{\text{small}} \ell_{\min}/\lambda \sim 1$, where $u_{\text{small}}$ is the characteristic velocity associated with $\ell_{\min}$. Equivalently, we might write $\lambda/\ell_{\min}^2 \sim \omega_{\text{small}}$, where $\omega_{\text{small}} \sim u_{\text{small}}/\ell_{\min}$ is a characteristic small-scale vorticity. If the estimate $\lambda/\ell_{\min}^2 \sim \omega_{\text{small}}$ is correct, then we have

$$\frac{1}{V_C} \left| \int_{R_C} \beta \overline{T'\mathbf{u}} \cdot \mathbf{g} dV \right| \sim \frac{B^2}{\rho\mu} \omega_{\text{small}}. \qquad (5)$$

[turbulent, low-$\Pr_m$ dynamo]

As we shall see, the energy production term on the left of (5) is proportional to the radial heat flux out of the core, and since the mantle provides the dominant thermal resistance to heat flow, that heat flux is independent of the dynamics of the core, and hence of $\Omega$. It follows that the left-hand side of (5) should be independent of the background rotation rate. We conclude, therefore, that the magnetic energy density does not depend on $\Omega$ provided that the magnitude of the small-scale vorticity is itself independent of $\Omega$, and indeed this is exactly what is suggested by the rotating convection experiments of Kunnen *et al*, 2010. It is the presence of the rate of working of the buoyancy forces and the apparent absence of $\Omega$ in (5) which provides independent motivation for the hypothesis that, in the saturated state, $B^2/\rho\mu$ is determined by the energy generation rate and does not exhibit any explicit dependence on the magnitude of the background rotation. Of course, these arguments are somewhat tenuous, yet when combined with the compelling numerical evidence that $B^2/\rho\mu$ is controlled only by the production of energy by buoyancy forces (Christensen, 2010), it provides some motivation to explore the consequences of the hypothesis that $B^2/\rho\mu$ is independent of the rotation rate.

We must, however, add one caveat to these statements. It has become common in some dynamo simulations to prescribe the temperature difference across the core, rather than prescribe the radial heat flux. In such a situation the heat flux, and hence the rate of working of the buoyancy forces, may be influenced by $\Omega$, and as a result there may be an implicit dependence of $B^2/\rho\mu$ on $\Omega$. Of course, such a dependence is simply an artefact of the boundary conditions used in those particular simulations. Never-the-less, from time to time we shall want to invoke the results of the numerical simulations in order to test our scaling laws. Consequently, when we refer below to $B^2/\rho\mu$ being independent of $\Omega$, we really mean that any influence of $\Omega$ on $B^2/\rho\mu$ occurs only though its indirect influence on the rate of working of the buoyancy forces, an influence that should vanish when fixed-flux boundary conditions are used.

In any event, according to the hypothesis above, we might expect $B^2/\rho\mu$ to be a function of the rate of production of energy by the buoyancy forces and, perhaps, of the scale of the convective flow, as represented by the integral scale of the motion, $\ell \sim R_C$. If P is the energy production rate per unit mass, then we have $B^2/\rho\mu = F(\ell, \mathrm{P})$, and dimensional analysis then demands

$$\frac{B^2}{\rho\mu} \sim \ell^{2/3} \mathrm{P}^{2/3} . \tag{6}$$

From (3) the time-averaged rate of production of energy per unit mass is $\mathrm{P} = -\beta\overline{T'\mathbf{u}} \cdot \mathbf{g}$, which includes contributions from both the steady-on-average convection and the turbulence. Moreover, the time-averaged convective heat flux per unit area, $\mathbf{q}_T$, is given by $\mathbf{q}_T/\rho c_p = \overline{T'\mathbf{u}}$, from which we conclude

$$\mathrm{P} = \frac{g\beta}{\rho c_p} q_T , \tag{7}$$

where $q_T$ is the radial heat flux through the core and $g = |\mathbf{g}|$. Thus the suggestion that $B^2/\rho\mu$ is a function of P and $\ell$ alone becomes

$$\frac{B^2}{\rho\mu} \sim \ell^{2/3} \left( \frac{g\beta q_T}{\rho c_p} \right)^{2/3} . \tag{8}$$

When $\mathrm{Pr}_m$ is not small, a significant fraction of the energy dissipation will occur through viscous stresses. In such cases (5) can be rewritten as

$$f(\mathrm{Pr}_m) \frac{1}{V_C} \left| \int_{R_C} \beta\overline{T'\mathbf{u}} \cdot \mathbf{g} dV \right| \sim \frac{B^2}{\rho\mu} \omega_{\mathrm{small}} ,$$

where $f$ is the fraction of the total dissipation that occurs as Ohmic dissipation. We would expect $f$ to be a monotonically decreasing function of $\mathrm{Pr}_m$, with $f = 1$ for $\mathrm{Pr}_m \ll 1$. Equation (8) then generalises to

$$\frac{B^2}{\rho\mu} \sim \ell^{2/3} \left( f \frac{g\beta q_T}{\rho c_p} \right)^{2/3} . \tag{9}$$

In this paper we restrict ourselves to low-$\Pr_m$ dynamos and so use (8) rather than (9). In principal the theory developed below could be generalised to include dynamos in which $\Pr_m$ is not small by multiplying P, or equivalently $q_T$, by $f(\Pr_m)$. However, since the functional form of $f(\Pr_m)$ has yet to be determined (it is almost certainly not a simple power law), and planetary cores invariably satisfy $\Pr_m \ll 1$, we shall leave aside most questions as to the role of $\Pr_m$ in those dynamos for which $\Pr_m$ is not small.

Note that, so far, we have not assumed any particular scaling for $u = |\mathbf{u}|$, as the logic above rests solely on the assumption that $B^2/\rho\mu$ is independent of $\Omega$ and is a function of P and $\ell$ alone. This is important because the scaling for $u$ has proven to be a subtle issue over which there is some disagreement. (Some of the scalings which have been proposed for $u$ are based on the assumption that $\mathbf{u}\cdot\nabla\mathbf{u}$ is an order-one quantity, which seems improbable for most planetary dynamos in the light of the extremely low value of Ro.) We shall return to the issue of how $u$ scales shortly.

If we make the tentative estimate $\ell \sim R_C$, and use the core-mantle (or outer core) heat flux, $q_C$, as a measure of $q_T$, we arrive at

$$\frac{B^2}{\mu} \sim \rho^{1/3} R_C^{2/3} \left(\frac{q_C}{c_P/g\beta}\right)^{2/3},$$

which is proposed in Christensen, Holzwarth & Reiners (2009) and Christensen (2010) based on a somewhat different line of reasoning. The scaling law $B^2 \sim q_C^{2/3}$ has been tested by Christensen (and others) against a wide range of numerical experiments and the results are surprisingly accurate over ranges of $B^2/\rho\mu$ and Ro that both span almost three-decades (*e.g.* $10^{-4} < \text{Ro} < 0.1$). There is also evidence that the scaling $B^2 \sim \rho^{1/3} q_C^{2/3}$ provides a consistent way of collapsing the data for the Earth and the giant planets (Christensen, 2010). So, all-in-all, there is considerable numerical support for (8).

Consequently, we might tentatively accept (8) as an empirically established relationship which can be rationalised through the observation that the small-scale vorticity in rapidly-rotating turbulent convection is independent of $\Omega$. We can then ask what other scaling laws follow from (8) when combined with the requirement for a force balance in the core. Let us, therefore, explore this force balance.

**3. The Core Force Balance and Resulting Scaling Laws**

We would expect the Coriolis, Lorentz and buoyancy forces all to be of similar magnitudes when the dynamo saturates. So, taking the curl of the momentum equation (2) to eliminate pressure, we have

$$\mathbf{\Omega}\cdot\nabla\mathbf{u} \sim \nabla\times(\beta T'\mathbf{g}) \sim \nabla\times(\mathbf{B}\cdot\nabla\mathbf{B}/\rho\mu). \qquad (10)$$

Note that we are obliged to omit the non-linear inertial term $\mathbf{u}\cdot\nabla\mathbf{u}$ from this force balance because the Rossby number is extremely small in planetary cores. Now we know that rapidly-rotating systems tend to develop elongated columnar vortices aligned with the rotation axis, so the minimum acceptable model must involve at least two length scales in the force balance. Consequently we let $\ell_{//}$ and $\ell_\perp$ be *integral length scales* parallel and perpendicular to the rotation axis, with $\ell_{//} \gg \ell_\perp$ and $\ell_{//} \sim R_C$. For large enough magnetic Reynolds number, $R_m = u\ell_{//}/\lambda$, we expect $\ell_\perp$ to be significantly larger than the micro-scale $\ell_{min}$ which appears in the energy balance (4), and so it is important to distinguish between $\ell_\perp$ and $\ell_{min}$. The force balance (10) now yields

$$\frac{\Omega u}{\ell_{//}} \sim \frac{g\beta T'}{\ell_\perp} \sim \frac{B^2}{\rho\mu\ell_\perp^2} \quad, \tag{11}$$

where we have assumed that the Lorentz force is dominated by the large-scale contributions to **B**. Next we combine (11) with the definition $P = -\beta\overline{T'\mathbf{u}}\cdot\mathbf{g} \sim g\beta T'u$, and with the low-$Pr_m$ estimate $B^2/\rho\mu \sim \ell_{//}^{2/3}P^{2/3}$, to give

$$\frac{\ell_\perp}{\ell_{//}} \sim \left(\frac{P}{\Omega^3\ell_{//}^2}\right)^{1/9}, \tag{12}$$

$$\frac{B^2/\rho\mu}{u^2} \sim \left(\frac{P}{\Omega^3\ell_{//}^2}\right)^{-2/9}, \tag{13}$$

$$R_m = \frac{u\ell_{//}}{\lambda} \sim \frac{\sigma B^2}{\rho\Omega}\left(\frac{P}{\Omega^3\ell_{//}^2}\right)^{-2/9} = \Lambda\left(\frac{P}{\Omega^3\ell_{//}^2}\right)^{-2/9}, \tag{14}$$

and

$$\text{Ro} = \frac{u}{\Omega\ell_{//}} \sim \left(\frac{P}{\Omega^3\ell_{//}^2}\right)^{4/9} = \left(\frac{(g\beta/\rho c_p)q_T}{\Omega^3\ell_{//}^2}\right)^{4/9}. \tag{15}$$

(Expressions (13) and (14) are simply rearrangements of (15) and (8), though (12) can be considered as an additional prediction.) We shall test (15) against the numerical simulations in §4, where we shall see that it is an excellent fit to the data.

Note that the diffusivities $\lambda$ and $\nu$ play no role in this scaling theory, except to the extent that $\lambda$ appears in the definition of $R_m$. This is, of course, typical of turbulent motion in which the diffusivities are small: the key dynamical processes are controlled by the large-scale dynamics which do not normally depend on the diffusivities, provided that those diffusivities are sufficiently small.

At this point it seems appropriate to return to (4) and ask if our proposed scaling laws are consistent with the low-$Pr_m$ energy balance

$$\int_{R_C}\beta\overline{T'\mathbf{u}}\cdot\mathbf{g}\,dV + \int_{R_C}\overline{\mathbf{J}^2}/\rho\sigma\,dV = 0. \tag{16}$$

Recall that $\ell_{min}$ is the smallest characteristic length scale associated with the magnetic field. Then (4) combined with the empirical law (8) demands

$$P \sim \frac{\mathbf{J}^2}{\rho\sigma} \sim \frac{B^2}{\rho\mu}\frac{\lambda}{\ell_{min}^2} \sim \ell_{//}^{2/3}P^{2/3}\frac{\lambda}{\ell_{min}^2}, \tag{17}$$

and hence

$$P^{1/3}\ell_{//}^{-2/3} \sim \frac{\lambda}{\ell_{min}^2} \sim \omega_{small}. \tag{18}$$

However (15) can be written as $P^{1/3} \sim \Omega \ell_{//}^{2/3} \text{Ro}^{3/4} = u\ell_{//}^{-1/3} \text{Ro}^{-1/4}$, or equivalently, $P^{1/3} \sim u\ell_{//}^{2/3}/\ell_\perp$, where we have used (12) to rewrite Ro in terms of $\ell_{//}/\ell_\perp$. Combining this with (18), we see that the low-$\text{Pr}_m$ energy balance (16) effectively imposes a constraint on $\ell_{\min}$:

$$\frac{\ell_{\min}^2}{\lambda} \sim \frac{\ell_\perp}{u}, \tag{19}$$

or equivalently,

$$\frac{\ell_{\min}}{\ell_\perp} \sim \left(\frac{u\ell_\perp}{\lambda}\right)^{-1/2}. \tag{20}$$

A number of interesting observations stem from (19) and (20). First, we note that $u/\ell_\perp$ is the characteristic large-scale vorticity associated with the columnar convection rolls, $\omega_{\text{large}} \sim u/\ell_\perp$. On the other hand, we know from §2 that $\lambda/\ell_{\min}^2$ is associated with the small-scale vorticity, $\lambda/\ell_{\min}^2 \sim \omega_{\text{small}}$. It follows from (19) that $\omega_{\text{large}} \sim \omega_{\text{small}}$, which is the hallmark of two-dimensional turbulence (Davidson, 2004) and might also be expected to hold in highly anisotropic turbulence in which $\ell_{//} \gg \ell_\perp$. Moreover, self-consistency of the model demands that we must extend the requirement that $\omega_{\text{small}}$ does not depend on the background rotation rate to the stronger constraint that $\omega_{\text{large}} \sim u/\ell_\perp$ is also independent of $\Omega$.

Second, following Christensen (2010) we might call $\tau_\lambda = \ell_{\min}^2/\lambda$ the Ohmic dissipation time, since equation (4) has $\ell_{\min}^2/\lambda$ proportional to the ratio of the magnetic energy to the Ohmic dissipation rate. Moreover, it is convenient to introduce two different dimensionless versions of the Ohmic dissipation time,

$$\tau_\lambda^* = \frac{\lambda \tau_\lambda}{\ell_{//}^2}, \quad \text{and} \quad \tau_\lambda^\Omega = \Omega \tau_\lambda, \tag{21}$$

which, according to (12) and (19), scale as

$$\tau_\lambda^* \sim \frac{\lambda \ell_\perp}{u\ell_{//}^2} = \frac{\lambda}{u\ell_{//}} \frac{\ell_\perp}{\ell_{//}} \sim \frac{\text{Ro}^{1/4}}{R_m}, \tag{22}$$

and

$$\tau_\lambda^\Omega \sim \frac{\Omega \ell_\perp}{u} = \frac{\Omega \ell_{//}}{u} \frac{\ell_\perp}{\ell_{//}} \sim \text{Ro}^{-3/4}. \tag{23}$$

Empirical scaling laws for the dependence of $\tau_\lambda^*$ and $\tau_\lambda^\Omega$ on Ro and $R_m$ have been determined by several authors, most recently by Christensen (2010), Yadav *et al* (2012) and Stelzer & Jackson (2013), based on analysing a multitude of numerical experiments. This provides yet another test of our predicted scaling laws.

Third, (20) highlights a potential difficulty which arises from the modest value of $R_m$ in the core of the terrestrial planets, and the consequent lack of a separation of scales. For example, in the Earth we have $u \sim 0.2\,\text{mm/s}$ and $\text{Ro} \sim 10^{-6}$. It follows that $R_m = u\ell_{//}/\lambda \sim 200$ and $\ell_{//}/\ell_\perp \sim 30$, at least according to (12), and so (20) yields $\ell_{\min}/\ell_\perp \sim (u\ell_\perp/\lambda)^{-1/2} \sim 0.4$. Evidently, in practice, it is not always easy to distinguish between the smaller of the two integral scales, $\ell_\perp$, and the micro-scale $\ell_{\min}$. Never-the-less, it is important to make the distinction from a conceptual point of view as $\ell_\perp$ and $\ell_{min}$ are predicted to scale in very different ways, at least for large $R_m$. Note also that, for the Earth, $u\ell_\perp/\lambda$ and $u\ell_{min}/\lambda$ are both predicted to have values close to unity, *i.e.* $u\ell_\perp/\lambda \sim 7$ and $u\ell_{min}/\lambda \sim 3$.

**4 Testing the Scaling Laws**

Of course, estimates (12)→(15), as well as (22) and (23), must be treated with caution, as there could be a multitude of length scales in the core, some associated with convection rolls, others with boundary layers, and yet others with internal shear layers and current sheets. In the scaling analysis above we simply scale all the forces on two integral scales, $\ell_\parallel \sim R_C$ and $\ell_\perp$, which we might associate with the axial and transverse dimensions of the columnar convection rolls. Never-the-less, we are free to ask if these scaling laws are consistent with what we know. The answer, by and large, is yes, at least to the extent that we can estimate the parameters involved.

Let us start by comparing our predictions with the published numerical experiments. In doing so it should be kept in mind that these simulations typically span the range $O(0.1) < \Pr_m < O(10)$, whereas we are concerned with the case of $\Pr_m \ll 1$. (For example, in the geodynamo, we have $\Pr_m \sim 10^{-6}$.) Thus the results of many of the numerical experiments exhibit a dependence on $\Pr_m$, a dependence which, according to §2, we would not expect to see at low $\Pr_m$. In any event, leaving aside the issue of the role of $\Pr_m$, a natural starting point for comparison is to consider prediction (15). This is consistent with the results of the numerical simulations compiled by Christensen & Aubert (2006) and Christensen (2010), which span almost three decades of Ro and show

$$\text{Ro} \sim \left(\frac{P}{\Omega^3 D^2}\right)^n, \tag{24}$$

where $D$ is the width of the spherical shell. For the particular numerical experiments included in Christensen (2010), the best-fit estimate of the exponent $n$ is $n = 0.41 \pm 0.03$, which is close to our prediction of $n = 0.444$. Moreover Yadav *et al* (2012), who compile and analyse the results of spherical dynamos with slip boundary conditions, suggest $n = 0.44$ in dipolar dynamos for both zonal and non-zonal measures of Ro, though interestingly the pre-factors in $\text{Ro}_{zonal} \sim \left(P/\Omega^3 D^2\right)^{0.44}$ and $\text{Ro}_{non\text{-}zonal} \sim \left(P/\Omega^3 D^2\right)^{0.44}$ are different. Finally, Stelzer & Jackson (2013), who re-examine and extend the data set of Christensen (2010), also estimate $n = 0.44$, all-be-it with a weak dependence on the magnetic Prandtl number (see below).

Figure 1 shows that data of Christensen & Aubert (2006) along with prediction (15). On the left is the full data set, corresponding to $0.06 < \Pr_m < 10$, while on the right only data in the range $0.06 < \Pr_m < 1$ is shown, which is closer to the low-$\Pr_m$ regime of interest here. It is clear that (15) is indeed a good fit to the data for $\Pr_m < 1$. So there is strong numerical support for our prediction of $n = 4/9$ in (24).

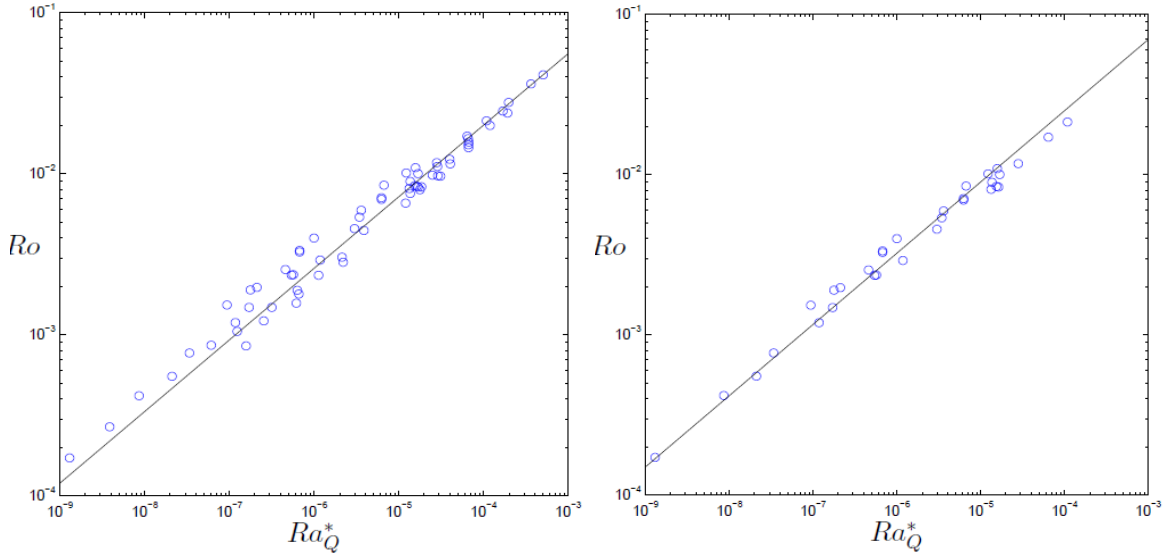

Figure 1. Computed values of Ro versus $Ra_Q^* = P/\Omega^3 D^2$ taken from the data set of Christensen & Aubert (2006). (a) The full data set, $0.06 < \Pr_m < 10$. (b) A restricted data set, $0.06 < \Pr_m < 1$. Power-law (15) is shown for comparison.

It is interesting to note in passing that Stelzer & Jackson (2013), who examine numerical experiments which span the wider range $0.06 < \Pr_m < 33$, suggest the empirical scaling

$$\text{Ro} \sim \left(\frac{P}{\Omega^3 \ell_{//}^2}\right)^{0.44} (\Pr_m)^{-\gamma}, \qquad (25)$$

where the exponent $\gamma$ is estimated to lie in the range $0.119 < \gamma < 0.133$, at least for the values of $\Pr_m$ considered. Rewriting this as

$$\text{Ro} \sim \left(f \frac{P}{\Omega^3 \ell_{//}^2}\right)^{0.44}, \qquad (26)$$

we see that $f$ in equation (9) is observed in Stelzer & Jackson (2013) to scale approximately as $f \sim (\Pr_m)^{-\beta}$, $0.27 < \beta < 0.30$. However, we require $f \to 1$ for $\Pr_m \ll 1$, and so we conclude that the expression $f \sim (\Pr_m)^{-\beta}$ can only be a local approximation to a more complex functional form for $f(\Pr_m)$. In fact we know this must be the case because extrapolating (25) down to the small values of $\Pr_m$ found in the core of the Earth significantly over-estimates Ro.

Next we consider estimates (22) and (23) for the dimensionless Ohmic dissipation time, which predict,

$$\tau_\lambda^* \sim \text{Ro}^{1/4}/R_m, \quad \tau_\lambda^\Omega \sim \text{Ro}^{-3/4}. \qquad (27)$$

These are close to the empirical observations of Yadav *et al* (2012), who estimate $\tau_\lambda^\Omega \sim \text{Ro}^{-0.8}$, and by implication $\tau_\lambda^* \sim \text{Ro}^{1/5}/R_m$, for numerical dynamos with slip boundary conditions. Moreover, (27) is close to the estimate of Christenson (2010) who, based on the evidence of many numerical experiments, suggests that $\tau_\lambda^*$ scales approximately as $\tau_\lambda^* \sim \text{Ro}^{1/6}/R_m$, which in turn implies $\tau_\lambda^\Omega \sim \text{Ro}^{-5/6}$.

The differences between the theoretical prediction, $\tau_\lambda^* \sim \mathrm{Ro}^{1/4}/R_m$, and the two empirical estimates, $\tau_\lambda^* \sim \mathrm{Ro}^{1/5}/R_m$ and $\tau_\lambda^* \sim \mathrm{Ro}^{1/6}/R_m$, are modest, differing as they do by factors of $\mathrm{Ro}^{1/20}$ and $\mathrm{Ro}^{1/12}$, respectively. Given that the variation in Ro is limited to three decades in the numerical experiments, and that the scatter in the data leaves some uncertainly in the exponent for Ro, one cannot readily use the numerical evidence to distinguish between the various proposed scaling relationships. Never-the-less, the fact that $\tau_\lambda^* \sim \mathrm{Ro}^{1/5}/R_m$ and $\tau_\lambda^* \sim \mathrm{Ro}^{1/6}/R_m$ are both reasonable fits to their respective numerical data sets does at least tell us that (27) must also be reasonably in line with that data.

Yet another estimate of $\tau_\lambda^*$, again based on the evidence of numerical simulations, is that of Stelzer & Jackson (2013) which, leaving aside any $\mathrm{Pr}_m$ dependence, suggests $\tau_\lambda^* \sim \mathrm{Ro}^\alpha/R_m$, $0.09 < \alpha < 0.11$. This is somewhat more at odds with our prediction of $\tau_\lambda^* \sim \mathrm{Ro}^{1/4}/R_m$, and this discrepancy remains to be accounted for. (The discrepancy possibly arises from the fact that the estimate $\tau_\lambda^* \sim \mathrm{Ro}^{0.1}/R_m$ is a curve fit over a relatively wide range of $\mathrm{Pr}_m$, specifically $0.06 < \mathrm{Pr}_m < 33$, whereas our prediction is for low $\mathrm{Pr}_m$ only.)

In addition to the evidence of the numerical simulations, we might note that (13) and (14) are consistent with what we know about the Earth, where we have $\mathrm{Ro} \sim 10^{-6}$ and $R_m \sim 200$ (see §3). For example, estimate (14) combined with (15) suggests $\Lambda \sim 0.2$, which is in line with table 2, where $\Lambda$ is estimated as $\Lambda \sim 0.08$. Also (13) combined with (15) suggests $(B^2/\rho\mu)/u^2 \sim 10^3$ which, given that $u \sim 0.2\,\mathrm{mm/s}$ in the core, suggests a mean core field strength of $B \sim 7\,\mathrm{Gauss}$. This is also in line with table 2 where the mean axial field strength based on the dipole moment **m** is estimated to be $B_z \sim 4\,\mathrm{Gauss}$.

In summary, then, predictions (15), (22) and (23) are all reasonably consistent with the evidence of the numerical experiments, while (13) and (14) are consistent with what we now about the geodynamo. The success of (15) lends support to (13) and (14), since these are simple rearrangements of (15) and (8).

## 5. Predictions for Other Planets

One of the difficulties of probing the consequences of scaling laws (12) → (15) for planets other than the Earth (and perhaps Jupiter) is that it is often difficult to obtain reliable estimates of the radial heat flux. However, one way forward is to note that there is reasonable numerical support for (8) and (15), as discussed above, at least for the type of dynamos generated by the numerical simulations. Let us, therefore, combine (8) with (15), while taking $\ell_{//} \sim R_C$ and $B \sim B_z$, where $B_z$ is the mean axial field strength in the core, as calculated from the dipole moment **m** in accordance with (1). This yields

$$\mathrm{Ro} \sim \left(\frac{B_z/\sqrt{\rho\mu}}{\Omega R_C}\right)^{4/3}, \qquad (28)$$

which allows us to estimate Ro, and hence $\mathrm{P}/\Omega^3 \ell_{//}^2$, in terms of quantities we can estimate with greater certainty, such as $\Omega$ and **m**. (Note that (28) may also be considered as a simple rearrangement of (14) and (15).)

Figure 2 shows that data of Christensen & Aubert (2006) for Ro versus $B_{rms}/\Omega D\sqrt{\rho\mu}$, along with prediction (28). ($D$ is the thickness of the fluid shell.) On the left is the full data set, corresponding to $0.06 < \mathrm{Pr}_m < 10$, while on the right only data in the range $0.06 < \mathrm{Pr}_m < 1$ is shown, which is closer to the regime of interest. Note that the characteristic field strength used by Christensen & Aubert (2006) is the rms magnetic field in the core, rather than the mean axial field calculated from (1). However, provided there is not a strong $\Omega$-effect, the two fields should be of similar magnitudes. It seems that (28) is not out of line with the numerical data, but the fit is somewhat disappointing when compared with figure 1. In some ways this is curious, since (28) simply combines (8) and (15), and figure 1 shows that (15) is an excellent fit to the data. We conclude that the scatter in figure 2 comes mostly from estimate (8). It is likely that the discrepancy arises from the use of data in which $\mathrm{Pr}_m \sim 1$, rather than $\mathrm{Pr}_m \ll 1$, which necessitates the use of (9) rather (8) to estimate the magnetic energy density. The problem, of course, is that we do not know the functional form of $f(\mathrm{Pr}_m)$ in (9).

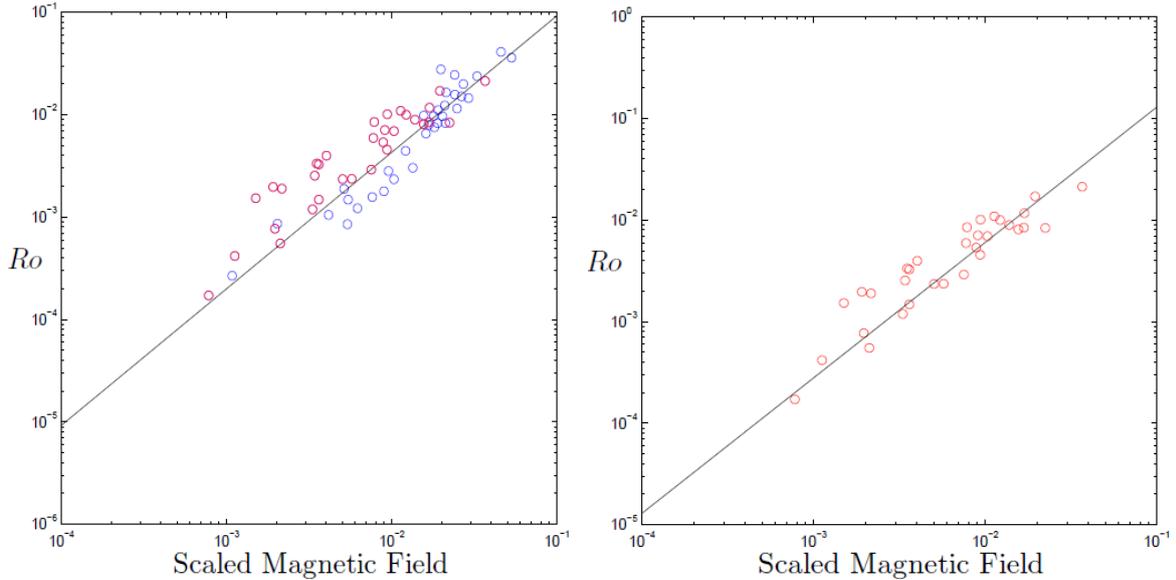

Figure 2. Computed values of Ro versus $B_{rms}/\Omega D\sqrt{\rho\mu}$ taken from the data set of Christensen & Aubert (2006). (a) The full data set, $0.06 < \mathrm{Pr}_m < 10$, with $\mathrm{Pr}_m < 1$ coloured red and $\mathrm{Pr}_m > 1$ coloured blue. (b) A restricted data set, $0.06 < \mathrm{Pr}_m < 1$. Power-law (28) is shown for comparison.

If one has sufficient confidence in power-law (28), it is tempting to use it to estimate Ro, and hence $P/\Omega^3 \ell_{//}^2$, in the core of the planets, although this is a rather substantial extrapolation. In any event, the various estimates of $B_z/\Omega R_C\sqrt{\rho\mu}$ are tabulated in table 2 and they are surprisingly consistent across the different planetary dynamos. Expression (28) then suggests Rossby numbers in the range $\mathrm{Ro} \sim 10^{-8} \to 10^{-6}$, as shown in table 3.

Given these estimates of Ro we can explore the predictions of scaling laws (12)→(14) for planets other than the Earth, though there is an implicit assumption here that the dynamos in these planets are qualitatively similar to those of the numerical simulations that provide support for (8) and (15). In any event, the tentative results of these scaling laws are shown in table 3. For simplicity, we have taken the pre-factors in (12), (13), (14), (15) and (28) to be unity, so little weight should be given to the precise numerical values. Three immediate observations are that: (i) for all of the planets, the predicted magnetic energy density in the core swamps the kinetic energy by a factor of $\sim 10^3$; (ii) the convection is predicted to be strongly anisotropic; and (iii), the magnetic Reynolds number for Saturn is surprisingly low, closer to that of the Earth than that of Jupiter. However, perhaps the primary observation is that Mercury appears not to fall into our scaling regime, as the predicted value of $R_m$ is much too low to sustain dynamo action. Once again, we are led to the conclusion that Mercury's dynamo is somehow anomalous.

| Planet | $\dfrac{B_z/\sqrt{\rho\mu}}{\Omega R_C}$ | $\text{Ro} = \dfrac{u}{\Omega R_C}$ | $\dfrac{B_z^2/\rho\mu}{u^2}$ | $\dfrac{\ell_{//}}{\ell_\perp}$ | $R_m \sim \dfrac{\Lambda}{\text{Ro}^{1/2}}$ |
|---|---|---|---|---|---|
| Mercury | $5.5\times 10^{-6}$ | $1\times 10^{-7}$ | $3\times 10^3$ | 60 | 0.2 |
| Earth | $1.3\times 10^{-5}$ | $3\times 10^{-7}$ | $2\times 10^3$ | 40 | 140 |
| Ganymede | $\sim 3.6\times 10^{-4}$ | $\sim 3\times 10^{-5}$ | $\sim 200$ | $\sim 10$ | $\sim 40$ |
| Jupiter | $5.2\times 10^{-6}$ | $1\times 10^{-7}$ | $3\times 10^3$ | 60 | 1600 |
| Saturn | $2.2\times 10^{-6}$ | $3\times 10^{-8}$ | $6\times 10^3$ | 80 | 130 |

Table 3. The order-of-magnitude predictions of scaling laws (12), (13) and (14) based on estimates of Ro given by (28). All pre-factors in these scaling relationships have been set to unity.

There is, however, a possible explanation for Mercury's apparently anomalous behaviour. The prediction $R_m \sim \Lambda/\text{Ro}^{1/2}$, combined with (1) and (28), leads to $R_m \propto R_C^{10/3}$, and so the predicted value of $R_m$ is extremely sensitive to the assumed core radius, or shell thickness if we take $\ell_{//} \sim D$. If, as suggested by Christensen, Schmitt & Rempel (2009), only the innermost regions of Mercury's core are in convective motion, then the effective values of $R_C$ or $D$ may be very much less than their nominal values. Moreover, $R_m \propto D^{10/3}$ tells us that reducing $D$ by a factor of, say, 4 would increase the predicted value of $R_m$ by a factor of 100. It remains to be seen as to whether or not this can account for Mercury's low value of $R_m$ in table 3.

## 6. Conclusions

We have developed new scaling laws based on Christensen's hypothesis that the saturated magnetic energy density should not depend on the planetary rotation rate. However, these scaling laws are different to those of Christensen (2010) and others because, in the light of the very small value of Ro in planetary cores, we insist that the non-linear inertial term, $\mathbf{u}\cdot\nabla\mathbf{u}$, is neglected in the core force balance. Our predictions are consistent with what we know about the Earth, at least to the extent that we can test them. Also, our predicted scalings for the Rossby number and dimensionless Ohmic dissipation time, $\text{Ro} \sim \left(P/\Omega^3\ell_{//}^2\right)^{4/9}$, $\tau_\lambda^* \sim \text{Ro}^{1/4}/R_m$ and $\tau_\lambda^\Omega \sim \text{Ro}^{-3/4}$, are reasonably consistent with the results of the numerical simulations shown in Christensen & Aubert (2006), Christensen (2010), Yadav *et al* (2012) and Stelzer & Jackson (2013). When applied to planets other than the Earth, our scalings suggest that Mercury's magnetic field is anomalous and that Mercury's dynamo either operates in a different regime, or is confined to a small part of the fluid core, as suggested by Christensen, Schmitt & Rempel (2009).

Acknowledgement: The author thanks Andy Jackson and Eric King for helpful comments and Andrea Maffioli for help in preparing the manuscript.